\documentclass{jjap3}

\title{Choice of Paper for Multigraphene Growth on Lead Pencil Drawing
}

\author{Satoru Kaneko$^{1,6}$, Yu Motoizumi$^{1}$, Kazuo Satoh$^{2}$, Yoshitada Shimizu$^{1}$, Manabu Yasui$^{1}$, Takeshi Rachi$^{1}$, Chihiro Kato$^{1}$, Satomi Tanaka$^{1}$, Mikio Ushiyama$^{3}$, Seiji Konuma$^{3}$, Yuko Itou$^{3}$, Tamio Endo$^{4}$, Akifumi Matsuda$^{5}$ and Mamoru Yoshimoto$^{5}$}

\inst{
$^{1}$Kanagawa Industrial Technology Research Institute,
Ebina, Kanagawa 243-0435 Japan\\
$^{2}$ Technology Research Institute of Osaka Prefecture, Izumi, Osaka 594-1157 Japan\\
$^{3}$ Kanagawa Academy of Science and Technology, Kawasaki, Kanagawa 213-0012 Japan\\
$^{4}$ Gifu University, Gifu 501-1193 Japan\\
$^{5}$ Tokyo Institute of Technology, Yokohama, Kanagawa 226-8503 Japan\\
}

\abst{
Graphene is prepared with a variety of methods, such as thermal decomposition of SiC, Chemical Vapor Deposition and pulsed laser deposition. We propose another method to prepare graphitic carbon by irradiating lead pencil drawn paper using femtosecond laser [Jpn. J. Appl. Phys. 55 (2016) 01AE24]. In order to evaluate electronic properties, such as resistivity and mobility, surface roughness of paper is important to evaluate thickness of lead pencil as one of principal factors. The surface roughness of variety of paper was investigated including cross section of drawing paper. Femotosecond laser irradiated the sample surface with its focus shifted from the optical focal point to enlarge the laser spot and tender irradiation.}

\begin{document}

\maketitle

\section{Introduction}

Graphene was first prepared by peeling scotch tape off graphite~\cite{novoselov2004,novoselov2005,sato2015}, and recently thermal decomposition of SiC~\cite{berger2004,hibino2008,bolen2009}, Chemical Vapor Deposition (CVD)~\cite{su2011,scott2011,jerng2011,michon2014} and pulsed laser deposition~\cite{cappelli2007,kumar2013} are popular methods to prepare graphene films. However, the thermal decomposition needs high temperature during the thermal process, and CVD requires transformation of film from metal catalyst to insulating substrates. We propose another method to prepare graphitic carbon by irradiating pencil drawn prints paper using femto-second laser. We drew a sheet of  printing paper with lead pencil, and irradiated the drawn paper by scanning femtosecond laser~\cite{kaneko2016a}.

In the previous study~\cite{kaneko2016a}, Raman method is mainly employed in order to evaluate quality of multigraphene grown on lead pencil drawing paper. However in order to measure other properties such as electronic resistivity and mobility, surface of lead pencil drawing is required to show flatness and thickness of the drawing area needs to be evaluated. Interestingly, pencil makers are interested in uniformity of drawing, which can be measured by reflection of visible light from the surface on lead pencil drawing area, not in the thickness of the drawing.

Although the surface of PC print paper looks like quite smooth on the surface, even optical microscope shows many of fiber on the surface, and the fiber on the surface prevent lead pencil from uniform and flat drawing on the paper. In this study, the surface roughness of  variety of paper were investigated by surface profiler and by taking cross section using scanning electron microscopy (SEM). Thickness of lead pencil drawing was estimated by the cross section images taken by SEM. Femtosecond laser was employed to irradiate the lead pencil drawing with blurred focused target point (focus shifted from optical focal point). 

Depending on ratio of content in pencil core, lead pencil was classified with the grade from H to B including F, as shown in Table~\ref{tbl:pencil_grade}, where ``H'' for hardness to ``B'' for blackness, and another grade ``F'' for fine. A set of pencils ranging from a very hard, light-marking pencil to a very soft, black-marking pencil usually ranges from hardest 10H to softest 10B with different content of carbon, clay and wax~\cite{sousa2001}. While the grade from 9H to 5H shows hallow peaks in Raman spectra, increasing carbon content resulted in increasing graphite peak (G peak) with less disorder peak  (D peak). Lead pencil used in this study was the grade of 10B pencil, which was not originally common grade of pencil but available at painting materials suppliers in recent.

\begin{figure}
\vspace{10 pt}
\center
\includegraphics[angle=0, scale=0.9]{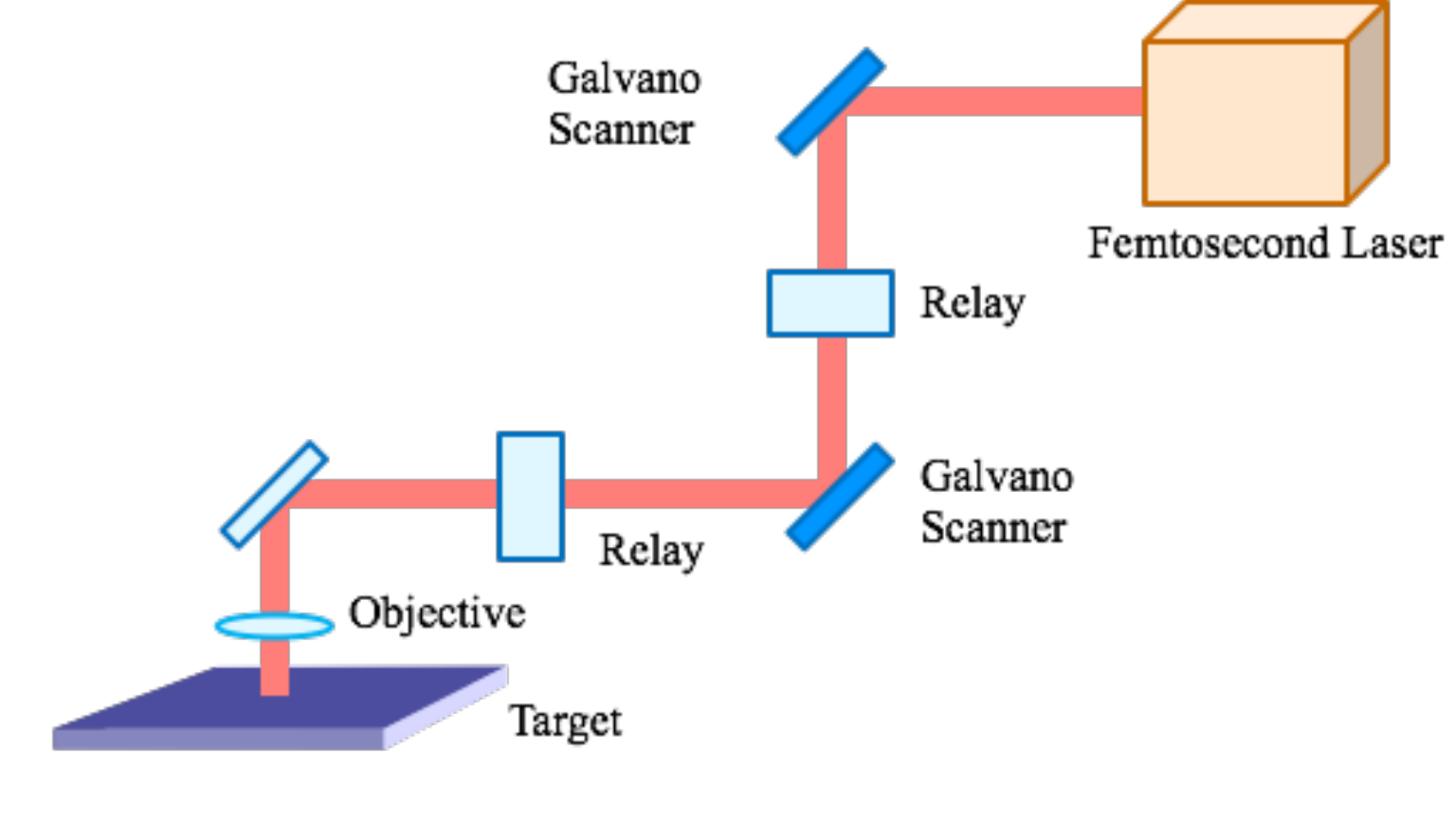}
\caption{\label{fig:Femtosec_Laser}(color online) Schematic of femtosecond laser with wavelength of 800 nm.}
\end{figure}

\begin{table*}[ht]
\begin{center}
\vspace{30 mm}
\caption{Grade of lead pencil. H (for hardness) to B (for blackness), as well as F (Fine), a letter arbitrarily chosen to indicate midway between HB and H.}
\label{tbl:pencil_grade}
\tiny
\begin{tabular}{ccccccccccccccccccccc}
Hardness & 9H & 8H & 7H & 6H & 5H & 4H & 3H & 2H & H & F & HB & B & 2B & 3B & 4B & 5B & 6B & 7B & 8B & 9B\\ \hline \hline
Carbon (\%) & 41 & 44 & 47 & 50 & 52 & 55 & 58 & 60 & 63 & 66 & 68 & 71 & 74 & 76 & 79 & 82 & 84 & 87 & 90 & 93\\ \hline
Clay (\%) & 53 & 50 & 47 & 45 & 42 & 39 & 36 & 34 & 31 & 28 & 26 & 23 & 20 & 18 & 15 & 12 & 10 & 7 & 5 & 2 \\ \hline
Wax (\%) & 5 & 5 & 5 & 5 & 5 & 5 & 5 & 5 & 5 & 5 & 5 & 5 & 5 & 5 & 5 & 5 & 5 & 5 & 5 & 5 \\
\end{tabular}
\end{center}
\end{table*}

\begin{figure*}[h!]
\center
\vspace{8mm}
\includegraphics[angle=0, scale=0.3]{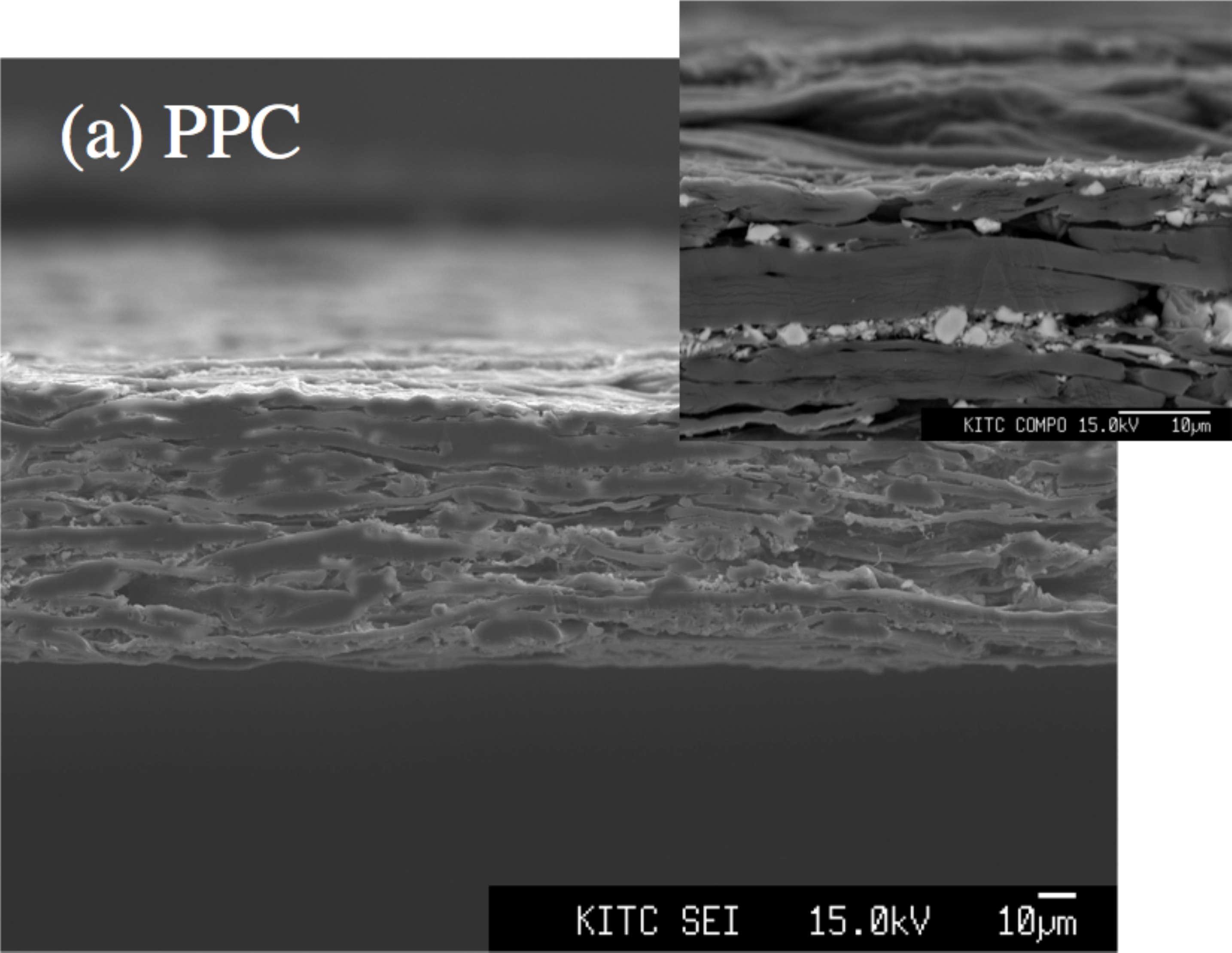}
\includegraphics[angle=0, scale=0.3]{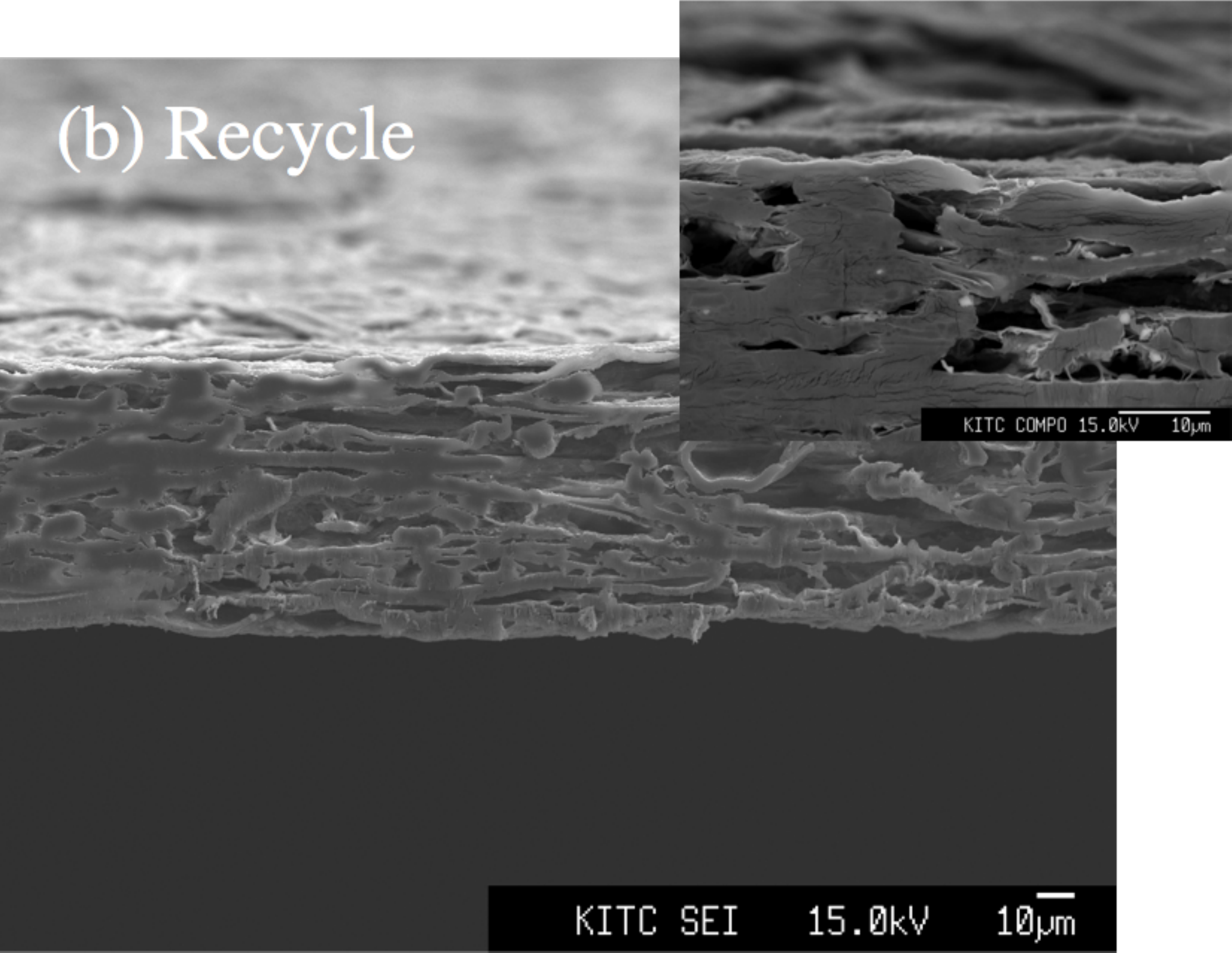}
\includegraphics[angle=0, scale=0.3]{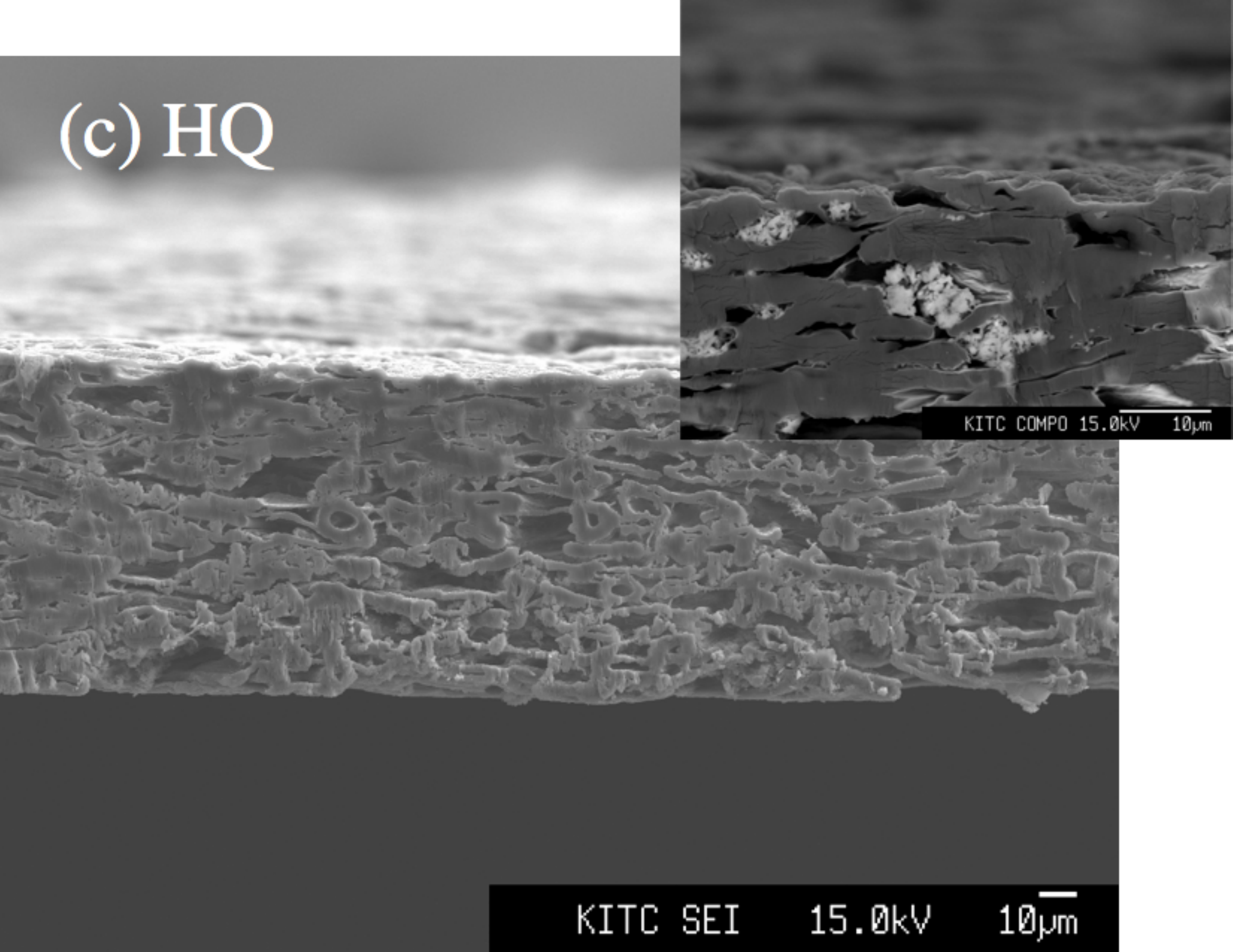}
\caption{\label{fig:cross_sections}Cross section images of (a) (PPC), (b) recycle paper and (c) high-quality paper (HQ). All the samples were prepared using a cross section polisher in dry environment without resin embedding.}
\end{figure*}

\section{Experimental}

\begin{table}[h]
\caption{Conditions for scanning femtosecond laser. The object indicates prints paper for PC printer (PC paper), photo prints paper for PC printer (PC photo paper), kent paper and paper for silver halide prints (sliver prints paper).}
\label{tbl:conditions}
\vspace{10pt}
\begin{center}
\begin{tabular}{cc}  \hline \hline
Object		& PPC\\ 
(backside)		& Kent paper\\
			& Recycle paper\\
			& HQ paper\\
			& Silver Prints Paper\\ \hline
Laser Frequency		& 1 kHz\\ \hline
Wavelength		& 800 nm \\ \hline
Power	&  0 $\sim$ 5 mW\\ \hline
Pencil	&  10B \\ \hline
\end{tabular}
\end{center}
\end{table}

Variety of papers prepared in this study were (1) plain paper copier (PPC), (2) recycled paper, (3) high-quality pure paper (HQ paper), (4) kent paper and (5) paper for silver halide prints (sliver prints paper). Surface profiler (Surf Coder, Kosaka Lab.) was employed to roughly measure the surface roughness on the papers. After cross  section polisher was used to prepared cross section of each paper, SEM was employed to observe the cross section of the papers.

After drawing on paper (roughly size of 10 $\times$ 20 mm) with the grade of 10B lead pencil, Raman spectroscopy with wavelength of 632.8 nm was employed to observe G, D and 2D Raman peaks, corresponding to $\sim$ 1,600, 1,350 and 2,700 cm$^{-1}$, respectively. SEM image of the cross section after lead pencil drawing was taken to estimate the thickness of lead pencil drawing. 

A femtosecond laser, (Cyber Laser IFRIT), was employed to irradiate the 10B lead pencil drawing area, as shown in figure~\ref{fig:Femtosec_Laser}. The power of laser was 3 mW with wavelength of 800 nm at repetition rate of 1 kHz. The spot size of laser was about 50 $\mu$m at the focused point. The laser was not focused on paper surface as a target point. Instead of focusing the paper surface, laser irradiated paper surface with the focus shifted 10 mm from its;optical focal point for uniform irradiation on the drawing area. The femtosecond laser scanned the lead pencil drawn paper with scanning speed at 1,000  $\mu$m/sec. 
A piece of paper fully covered with graphite by pencil drawing was placed on X-Y stage, and the femtosecaond laser was scanned with a line pitch of 25 $\mu$m. The details of irradiation conditions were shown in Table~\ref{tbl:conditions}.
Raman was again employed after the laser irradiation on paper surface.

\section{Results \& Discussions}

\begin{table}
\caption{Summary of surface roughness on Plain Paper Copier (PPC), Kent paper, Recycled paper and Print Paper. Ra and Rz indicate arithmetical mean roughness and ten-point mean roughness, respectively}
\begin{center}
\begin{tabular}{ccc}
Paper & R$_{a}$ ($\mu$m) & R$_{z}$ ($\mu$m)\\ \hline \hline
PPC & 2.95 & 15.44 \\
Kent paper & 2.57 & 15.21 \\
Recycle paper & 1.97 & 11.09 \\
HQ paper & 1.76 & 11.49 \\
Print Paper & 1.25 & 6.91 \\
\end{tabular}
\end{center}
\label{tbl:roughness}
\end{table}%

Surface profilometer was used to evaluate the surface roughness of variety of papers, as shown in Table~\ref{tbl:roughness}. While PPC and Kent paper showed roughness of R$_{a}$ 2 $\sim$ 3 and R$_{z}$ $\sim$ 15 $\mu$m, the surface roughness R$_{a}$ of print paper was about 1 $\mu$m and R$_{z}$ about 7 $\mu$m. Although the roughness of print paper was only a half of ones of PPC and kent papers, the surface of papers were quite different. PPC and kent papers including recycle and HQ papers, showed many of fiber structure on the surface compared to one of print paper.

The details of paper surface were observed by SEM images of cross section samples prepared using cross section polisher produced by JEOL. The cross section polisher uses an argon beam to mill cross sections or polish virtually materials with the continuously rotating sample holder, which prepare cross section sample in dry environment without sample embedding in resin. Figure~\ref{fig:cross_sections} shows cross section images of PPC and HQ paper prepared using cross section polisher without resin embedding. The SEM images clearly show fiber structure consisting of cellulose together with fillers as a loading material (white particles in cellulose)~\cite{motoizumi2013}.
%
White particles in the SEM images are calcium carbonate, which is often used as loading materials in many kinds of papers. Interestingly, high quality paper uses natural growth of calcium carbonate consisting of fine small particles, as shown in the inset of Fig.~\ref{fig:cross_sections}. PPC uses artificial materials, of which size is relatively larger than natural materials and artificial calcium carbonate usually shows round shape while natural one shows rugged and angular shape~\cite{motoizumi2013}. The round and large size of loading materials are shaken down at roller in printer, which often causes running idle of paper feeding. Many papers showed roughness R$_{a}$ of 2 $\sim$ 3 $\mu$m with fiber texture on the surface.

\begin{figure}
\center
\includegraphics[angle=0, scale=0.6]{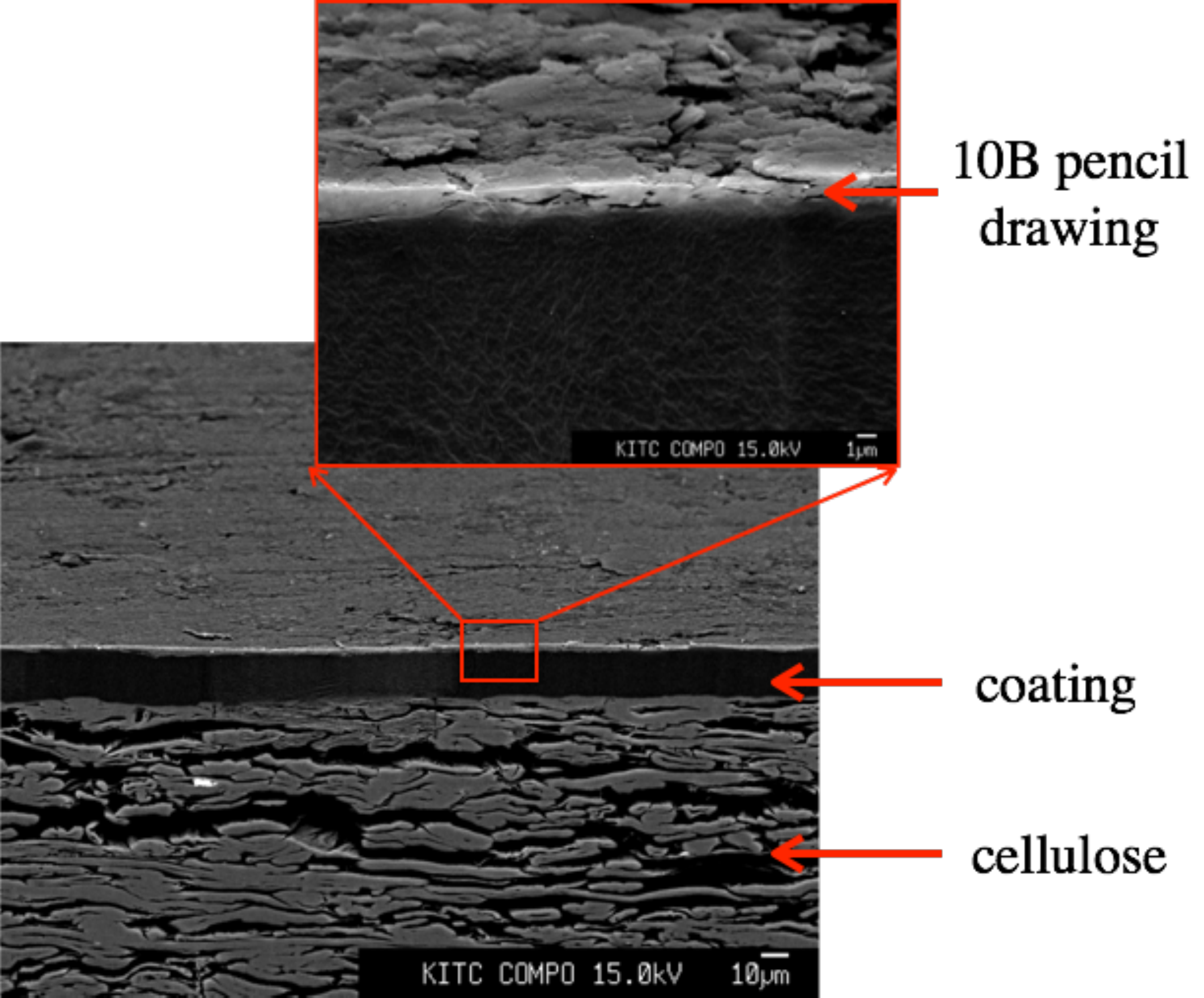}
\caption{\label{fig:cross_section_print_paper}Cross section images of (a) (PPC), (b) recycle paper and (c) high-quality paper (HQ).}
\end{figure}

An inkjet paper has flat and smooth surface with variety of coating. Matte inkjet papers, for example, is common to use silica as pigment together with polyvinyl alcohol (PVOH), and glossy inkjet papers can be made by multicoating, resin coating, or cast coating on a lamination paper. However, lead pencil can hardly draw on such smooth paper. The coating on inkjet paper is not suitable for pencil drawing.
Only silver-halide print paper in the Table~\ref{tbl:roughness} shows relatively flat surface without fiber structure, and lead pencil can draw on the back of print paper.
Figure~\ref{fig:cross_section_print_paper} shows a cross section of print paper after 10B lead pencil drawing. Interestingly, a thin resin coating layer was observed on layer of cellulose even on back of print paper, and pencil drawing layer of $\sim$ 2 $\mu$m is clearly observed on the resin coating layer.

\begin{figure}
\center
\includegraphics[angle=0, scale=0.7]{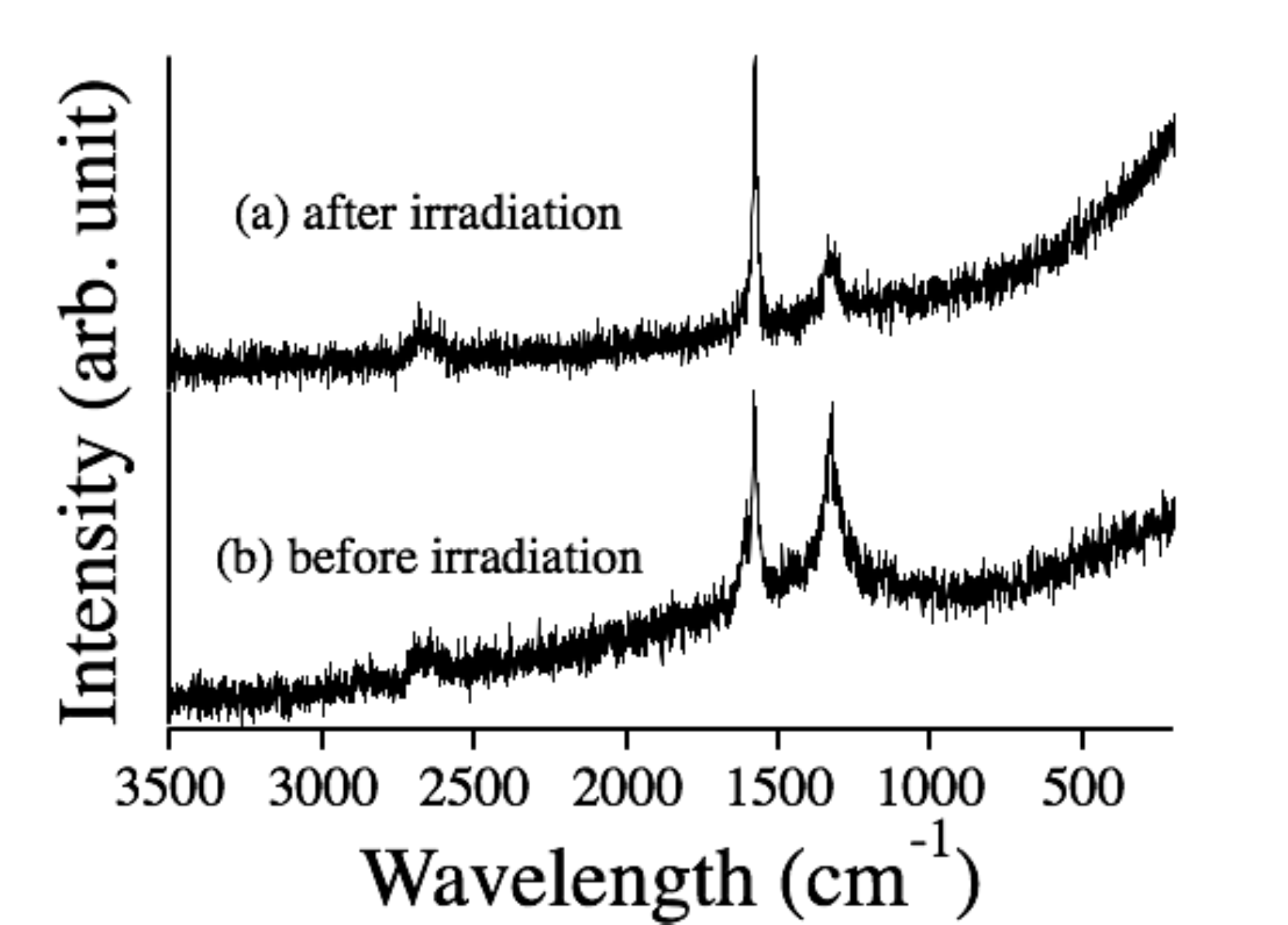}
\caption{\label{fig:raman}.Raman spectra from lead pencil drawing on print paper (a) after and (b) before laser irradiation.}
\end{figure}

10B lead pencil drawing on print paper was irradiated using femtosecond laser with focus shifted from the paper surface (10mm along z-axis). Raman spectroscopy revealed strong G peak and less intensity of D peak, as shown in Fig.~\ref{fig:raman}, and the thickness of lead pencil drawing became less than half after the laser irradiation.
For electronic properties such as resistivity and mobility, optimal condition of irradiation, such as laser power, focus shift, scanning speed are required.

In summary, variety of papers were evaluated for multigraphene growth on lead pencil drawing. The thickness of pencil drawing was estimated on print paper by taking SEM images of cross section of variety of papers prepared using cross section polisher. Cross section polisher is useful method to prepare a cross section of a variety paper for taking SEM images. Femtosecond laser was employed to irradiate the lead pencil drawing with shifted focal point to expand unifomity of laser irradiation on the target surface. Raman showed G' peak with less intensity of D peak, after the irradiation, however optimal conditions for laser irradiation is required to evaluate other properties such as resistivity and mobility.

In summary, a variety of paper was drawn by lead pencil with the grade between 4H through 10B. Raman spectroscopy verified both G and D peaks, corresponding sp$^{2}$ hybridisation (graphitic signature of carbon) and disorder due to defects. After irradiation of scanning femtosecond laser on silver paper drawn by 10B lead pencil, Raman spectroscopy showed the laser scanning eliminated D peak and remained G and 2D peaks, which indicating the graphene remaining on silver prints paper.

\acknowledgement
We would like to thank Toru Katakura at SONY Corp. and Yoshitsugu Sato at Kanagawa Industrial Technology Center for technical support. This research was supported in part by Grant- in-Aid for Scientific Research (KAKENHI) C 26420692.

\end{document}